\documentclass[aps,showpacs,prl,twocolumn,superscriptaddress,footinbib]{revtex4-1}
\usepackage{epsfig}
\usepackage{amsfonts}
\usepackage{amsmath}
\usepackage{amssymb}
\usepackage{bm}
\usepackage[dvipsnames]{xcolor}
\usepackage{booktabs}
\usepackage{yfonts}

\usepackage[utf8]{inputenc}
\usepackage[colorlinks,bookmarks=false,citecolor=blue,linkcolor=blue,urlcolor=blue]{hyperref}
\usepackage{soul}

\usepackage{changepage}

\newcommand{\ket}[1]{\left| #1 \right>}
\newcommand{\braket}[2]{\left< #1 | #2 \right>}

\newcommand{\av}[1]{\left< #1 \right>}

\newcommand{\be}{\begin{equation}}
\newcommand{\ee}{\end{equation}}

\newcommand{\er}[1]{Eq.~\eqref{#1}}

\begin{document}

\title{Classical Stochastic Discrete Time Crystals}
\author{F. M. Gambetta}
\affiliation{School of Physics and Astronomy, University of Nottingham, Nottingham, NG7 2RD, United Kingdom and Centre for the Mathematics and Theoretical Physics of Quantum Non-equilibrium Systems, University of Nottingham, Nottingham NG7 2RD, UK}
\author{F. Carollo}
\affiliation{School of Physics and Astronomy, University of Nottingham, Nottingham, NG7 2RD, United Kingdom and Centre for the Mathematics and Theoretical Physics of Quantum Non-equilibrium Systems, University of Nottingham, Nottingham NG7 2RD, UK}
\author{A. Lazarides}
\affiliation{Interdisciplinary Centre for Mathematical Modelling and Department of Mathematical  Sciences, Loughborough  University, Loughborough LE11 3TU,  UK}
\author{I. Lesanovsky}
\affiliation{School of Physics and Astronomy, University of Nottingham, Nottingham, NG7 2RD, United Kingdom and Centre for the Mathematics and Theoretical Physics of Quantum Non-equilibrium Systems, University of Nottingham, Nottingham NG7 2RD, UK}
\affiliation{Institut für Theoretische Physik, Universität Tübingen, Tübingen 72076, Germany}
\author{J. P. Garrahan}
\affiliation{School of Physics and Astronomy, University of Nottingham, Nottingham, NG7 2RD, United Kingdom and Centre for the Mathematics and Theoretical Physics of Quantum Non-equilibrium Systems, University of Nottingham, Nottingham NG7 2RD, UK}

\date{\today}

\begin{abstract}
We describe a general and simple paradigm for discrete time crystals (DTCs), systems with a stable subharmonic response to an external driving field, in a classical thermal setting. We consider, specifically, an Ising model in two dimensions, as a prototypical system with a phase transition into stable phases distinguished by a local order parameter, driven by a thermal dynamics and periodically kicked with a noisy protocol. By means of extensive numerical simulations for large sizes - allowed by the classical nature of our model - we show that the system features a true disorder-DTC order phase transition as a function of the noise strength, with a robust DTC phase extending over a wide parameter range. We demonstrate that, when the dynamics is observed stroboscopically, the phase transition to the DTC state appears to be in the equilibrium 2D Ising universality class. However, we explicitly show that the DTC is a genuine non-equilibrium state. More generally, we speculate that systems with thermal phase transitions to multiple competing phases can give rise to DTCs when appropriately driven. 
\end{abstract}

\maketitle

\noindent
{\bf \em Introduction.--} Discrete time crystals (DTCs)~\cite{Sacha:2015,Khemani:2016prl,Else:2016prl,vonKeyserlingk:2016,Khemani:2017,Yao:2017,Huang:2018,Choi:2017,Zhang:2017,Ho:2017,Else:2016prl,Else:2017prx,Russomanno:2017,Rovny:2018prl,Rovny:2018,Pal:2018,Yu:2019,Barfknecht:2018,Huang:2018,Flicker:2018,Yao:2018,Heugel:2019} have recently emerged as a novel form of non-equilibrium quantum matter. A time crystal is a system in which the time translation invariance of the dynamics is spontaneously broken asymptotically~\cite{Wilczek:2012,Shapere:2012}. While time crystallinity is impossible in thermal equilibrium states of systems with a local time-independent Hamiltonian~\cite{Bruno:2013,Nozieres:2013,Watanabe:2015,Kozin:2019}, it can emerge in excited states~\cite{Syrwid:2017} or through spontaneous breaking of an underlying discrete time symmetry~\cite{Sacha:2015,Else:2016prl,Khemani:2016prl}. Concrete examples of the latter mechanism have been found in closed quantum systems with time-periodic Hamiltonians (``Floquet'' systems~\cite{Shirley:1965,Sambe:1973,Grifoni:1998}) such as the $\pi$-spin glass and related models~\cite{Else:2016prl,Khemani:2016prl,vonKeyserlingk:2016,Khemani:2017,Yao:2017,Choi:2017,Zhang:2017,Ho:2017}. The usual setting in such unitary spin systems is that the dynamics is split between evolution with an interacting disordered Hamiltonian followed by a rotation of the spins. The usual Floquet heating towards an infinite temperature state~\cite{DAlessio2014,Lazarides:2014,Ponte2015} is avoided by exploiting localization~\cite{Lazarides:2015prl,Ponte2015b} so that spatio-temporal order can be established without fine-tuning. For reviews see, e.g., Refs.~\cite{Moessner:2017,Sacha:2018, Else:2019}.

 \begin{figure}[b]
	\centering
	\includegraphics[width=\columnwidth]{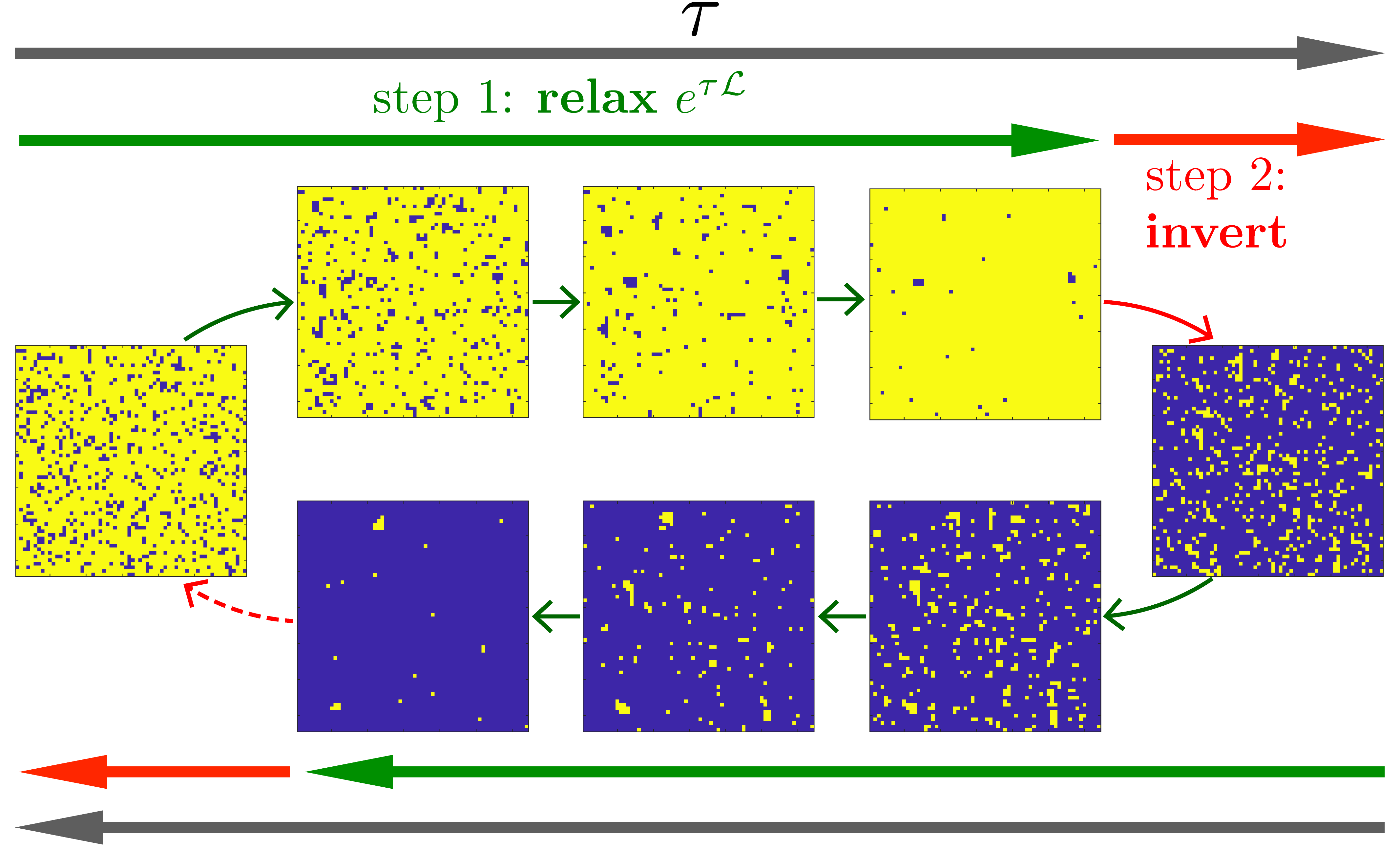}
	\caption{{\bf Dynamics of the driven Ising model under DTC conditions.}
	Starting a cycle from a magnetized state, domains of the opposite magnetisation (blue-dark/yellow-light indicate down/up, respectively) are annealed away under the evolution of duration $\tau$ generated by $\mathcal{L}$ during step 1 of the protocol. Step 2 randomly inverts most of the spins, leaving on average $\varepsilon=1-\mu$ ``mistakes''. These are subsequently annealed away in step 1 of the succeeding period. Due to the underlying symmetry breaking of the Ising model, only after two periods the original orientation of the spin profile is recovered. This gives rise to the period-doubling DTC. }
	\label{fig:sketch}
\end{figure}

The discovery of DTCs in unitary disordered Floquet systems raises numerous questions. Two important ones are the possibility of realizing DTCs in clean quantum systems \cite{Sacha:2015,Else:2017prx,Russomanno:2017,Yu:2019,Huang:2018,Barfknecht:2018,Giergiel:2018,Schafer:2019} and whether DTCs can survive in the presence of dissipation~\cite{Lazarides:2017,Iemini:2017,Wang:2018,Gong:2018,Gambetta:2019,Pal:2018,OSullivan:2018,Tucker:2018,Zhu:2019,Lazarides:2019,Droenner:2019,Buca:2019,Strinati:2019,Carlo:2019,Bello:2019}. Studying these two issues is part of the more general search for an understanding of the range of mechanisms through which time crystalline order can be stabilized. In the case of quantum systems coupled to an external environment, the problem one faces is that of the natural tendency of dissipation to destabilize order~\cite{Lazarides:2017, Zhang:2017,Choi:2017,Lazarides:2019}. In this respect, several mean-field or fully connected model systems have been shown to display DTC behavior with an appropriate engineering of the dissipative processes~\cite{Wang:2018,Gong:2018,OSullivan:2018,Droenner:2019,Cosme:2019,Kessler:2019,LLedo:2019}, with some candidate open quantum systems argued to do the same away from mean-field~\cite{Gambetta:2019,Lazarides:2019,Zhu:2019}. An important open question concerning the emergence of DTCs in open settings is their survival in the presence of thermal noise~\cite{Yao:2018,Heugel:2019,Oberreiter:2019,Lazarides:2019}. In this context, the emergence of DTCs in classical systems coupled to an external environment, inducing both damping and stochastic forces, have been recently investigated in Refs.~\cite{Yao:2018,Heugel:2019}. Interestingly, in Ref.~\cite{Yao:2018} an activated first-order dynamical DTC phase transition has been found in a one-dimensional driven Frenkel-Kontorova model at finite temperature while, in Ref.~\cite{Heugel:2019}, the stabilization of truly many-body and robust period-doubled states have been demonstrated in a network of coupled non-linear oscillators.

In this work, we shed light on the interplay between driving, noise, and interactions and the associated disorder-DTC order phase transition by considering a prototypical setting for DTCs in fully classical and thermal many-body systems. The generic scheme we propose is extremely simple, but to our knowledge has not been presented elsewhere before. The setup is that of a classical system with an equilibrium symmetry breaking transition - below we consider a two-dimensional (2D) Ising model as an obvious example - which is periodically driven. A period of the dynamics of duration $\tau$ consists of stochastic evolution under conditions such that asymptotically (i.e., if $\tau$ was to diverge) the symmetry-broken state would be stable, followed by a sudden random inversion of a fraction $\mu$ of the spins, see Fig.~\ref{fig:sketch}. We show that there exists a wide range of values of $(\tau,\mu)$ in which a stable DTC emerges. This driven dynamics, when observed stroboscopically, leads to a non-equilibrium stationary state (NESS), in terms of which we construct the associated phase diagram as function of $(\tau,\mu)$. We then show that, despite the out-of-equilibrium nature of the dynamics, the transition from disorder to DTC order appears to be in the 2D Ising universality class. 

\begin{figure}[t]
	\centering
	\includegraphics[width=\columnwidth]{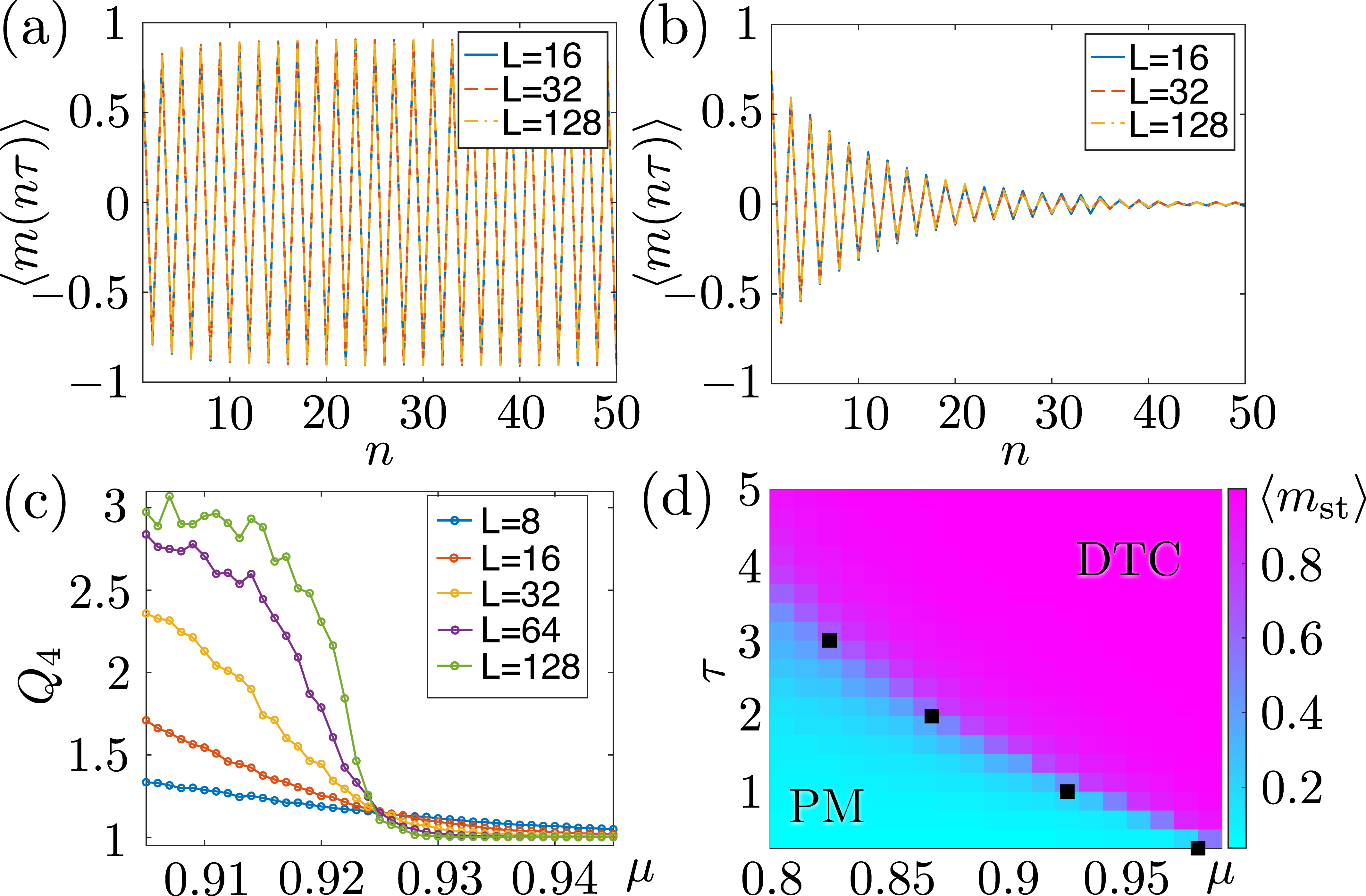}
	\caption{{\bf DTC phase transition.} (a) Magnetization $m(n \tau)$ at the end of each cycle $n$ for $(\tau,\mu)=(1,0.95)$ from an initial state with $m_{\rm in} = 0.8$. The persistent oscillations with period $2 \tau$ are indicative of a DTC. (b) Same for $(\tau,\mu)=(1,0.85)$. Here the oscillations get rapidly attenuated, indicative of the disordered time-homogeneous phase. (c) Binder cumulant $ Q_4 $ as a function of $ \mu $ for $\tau=1$ and different system sizes. The crossing of the curves locates the critical point of the disorder-DTC order phase transition at $\mu_c\approx0.925$. (d) Phase diagram as a function of $ \mu $ and $ \tau $. The shading corresponds to the value of $\av{m_{\rm st}}$ in a system with $L=32$. Black squares give the location of $\mu_c(\tau)$ extracted from the analysis of $ Q_4 $ at $\tau=\{0.25, 1, 2, 3\}$, cf.\ panel (c). }
	\label{fig:magandPD}
\end{figure}

\bigskip
\noindent
{\bf \em Model and dynamical protocol.--} 
We study a classical Ising model on a 2D square lattice of linear size $L$ and periodic boundary conditions, with nearest neighbour couplings and no magnetic field. The energy function is 
\begin{equation}\label{eq:E}
E({\bm{\sigma}})
\equiv
-J\sum_{\av{i,j}}\sigma_i \sigma_j,
\end{equation}
where $\sigma_i = \pm 1$, sites are labelled by $i,j=\{1,...,L^2\}$, $\av{i,j}$ denotes nearest neighbours, ${\bm{\sigma}} = \{ \sigma_i, \ldots, \sigma_{\mathcal{N}} \}$ with $\mathcal{N}=L^2$ indicates a whole configuration of spins, and the coupling is ferromagnetic, i.e., $ J>0 $. Below we set $J=1$ for simplicity.

The dynamics is periodic~\cite{Note1} according to the following two-step protocol within each period $ \tau $, see Fig.~\ref{fig:sketch}:
\begin{enumerate}
	\item The first step evolves the system for time $\tau$ with a single spin-flip thermal dynamics which obeys detailed balance with respect to the equilibrium Boltzmann distribution associated to the energy function of \er{eq:E} at temperature $T$. As we are interested in dynamics where a  symmetry broken state is stable, we will restrict to $T=0$ below. 
	\item The second step instantaneously inverts each spin with probability $\mu$. When $\mu < 1$ not all spins are inverted, and we denote with $\varepsilon=1-\mu$ the ``error'' introduced in the inversion step.
\end{enumerate}
This driven dynamics is described, within a period, by the evolution operator
\begin{equation}
\label{eq:Floquet}
\mathcal{F}_\tau 
\equiv
 \left[
\prod_{i=1}^{\mathcal{N}}\left( \varepsilon \, \mathbb{I}_i +\mu \, \hat{\sigma}^x_i \right) 
\right]
e^{\mathcal{L} \tau}, 
\end{equation}
with $ \hat{\sigma}^x_i $ and $ \mathbb{I}_i $ the $ x $ Pauli and the identity matrices acting on the $i$-th spin, respectively. The rightmost factor implements step 1 above, with
\begin{equation}\label{eq:Glauber}
\mathcal{L} 
\equiv
\sum_{i=1}^{\mathcal{N}}(\hat{\sigma}^x_i-\mathbb{I}_i) \, \Gamma_i({\bm{\sigma}})
\end{equation}
the generator of zero-temperature Glauber dynamics \cite{Glauber:1963,Binder:2010,Krapivsky:2010}. Here, $\Gamma_i({\bm{\sigma}})=\theta[E({\bm{\sigma}})-E({\bm{\sigma}}_i)]$, where ${\bm{\sigma}}_i$ indicates the configuration obtained from ${\bm{\sigma}}$ by flipping the $i$-th spin, and $\theta(z)$ the Heaviside function. Therefore, only moves that do not increase the energy are allowed. The leftmost factor in \er{eq:Floquet} implements step 2 above. 

After a transient, the system reaches a time-periodic NESS. The dynamics depends on two parameters, the period $\tau$ and the rotated fraction of spins $ \mu $. To investigate the DTC phase transition it is convenient to consider the {\em stroboscopic-time staggered} magnetization, 
\begin{equation}\label{eq:staggeredmag}
m_{\rm st}(n) 
\equiv
(-1)^{n} \, m(n \tau),  \qquad\text{with }  n={0,1,\ldots}
\end{equation}
where $m(t)$ denotes the total magnetization per spin at time $t$,
\begin{equation}\label{eq:mag}
m(t) 
\equiv
\frac{1}{\mathcal{N}}\sum_{i=1}^{\mathcal{N}} \sigma_i(t).
\end{equation}
For a large enough number of cycles, we might expect that a stroboscopic NESS will be established. To characterize it, a natural order parameter is the average of the stroboscopic-time staggered magnetization of \er{eq:staggeredmag}, 
\begin{equation}\label{eq:mstav}
\av{m_{\rm st}} = \lim_{N \to \infty} \frac{1}{N} \sum_{n=1}^{N} m_{\rm st}(n) .
\end{equation}
Note that $\av{m_{\rm st}}$ corresponds to the usual order parameter used to identify DTC order, given by the Fourier component of the magnetization at half the driving frequency~\cite{Moessner:2017,Sacha:2018,Else:2019}.  

\bigskip
\noindent
{\bf \em DTC state and phase transition.--} We numerically investigate the dynamics given in Eq.~\eqref{eq:Floquet} by using a continuous-time Monte Carlo algorithm~\cite{Bortz:1975,Newman:1999,Binder:2010} for the thermal part of the protocol. In Figs.~\ref{fig:magandPD}(a,b) we show the (stroboscopic) behavior of the magnetization at times which are multiples of the period, $m(n \tau)$, with $n = 0,1,2,\ldots$, for $\tau = 1$ and two different values of $\mu$. Two phases can be clearly identified: Fig.~\ref{fig:magandPD}(a) shows $m(n \tau)$ for $\mu = 0.95 > \mu_c $. Here, the system displays stable DTC oscillations with twice the period of the driving (ordered DTC phase). For the value of $ \tau $ considered in Fig.~\ref{fig:magandPD}, our estimate for the transition point between the ordered DTC and the disordered driven paramagnet is $\mu_c\approx0.925$ (see below for details). For a smaller value of $\mu = 0.85 < \mu_c$ (i.e., for a larger number of mistakes in the inversion) the oscillations in the magnetization decay quickly with time, see Fig.~\ref{fig:magandPD}(b), corresponding to the disordered (paramagnetic) phase (PM).

In terms of the staggered magnetization of Eq.~\eqref{eq:staggeredmag}, the disordered phase has $\av{m_{\rm st}}=0$ while in the DTC phase $\av{m_{\rm st}} \neq 0$. This suggests that a phase transition takes place at the transition point $ \mu_c $. To determine its nature and to precisely locate the transition point, we then inspect the behavior of Binder cumulants of the staggered magnetization~\cite{Newman:1999,Binder:2010,Sandvik:2010}
\begin{equation}\label{eq:Binder}
Q_{2p}=\frac{\av{m_{\rm st}^{2p}}}{\av{|m_{\rm st}|^p}^2},
\end{equation}
with $p \geq 2$. As shown in Fig.~\ref{fig:magandPD}(c), the curves for $ Q_{2p} $ as a function of $ \mu $ for different system sizes $ L $ and fixed $ \tau $ cross at a single point. The size-independence of $ Q_{2p} $ at this point implies the emergence of a diverging correlation length $ \xi $ (see below for details) associated with a continuous phase transition. Thus, the crossing point allows for a precise location of the critical point $ \mu_c $ for a given value of $ \tau $.  
The corresponding phase diagram of the disorder-DTC order phase transition in the $(\tau,\mu)$ plane can be obtained by repeating the procedure described in Fig.~\ref{fig:magandPD}(c) for different values of $ \tau $ and it is shown in Fig.~\ref{fig:magandPD}(d). Here, for a given $\tau$, there is a $\mu_c(\tau)$ such that $\mu < \mu_c(\tau)$ corresponds to the disordered phase, while $\mu > \mu_c(\tau)$ to the DTC phase.  $\mu_c(\tau)$ decreases with increasing $\tau$ as the longer annealing time allows for a larger error density before the DTC order becomes unstable.

\begin{figure}[t]
	\centering
	\includegraphics[width=\columnwidth]{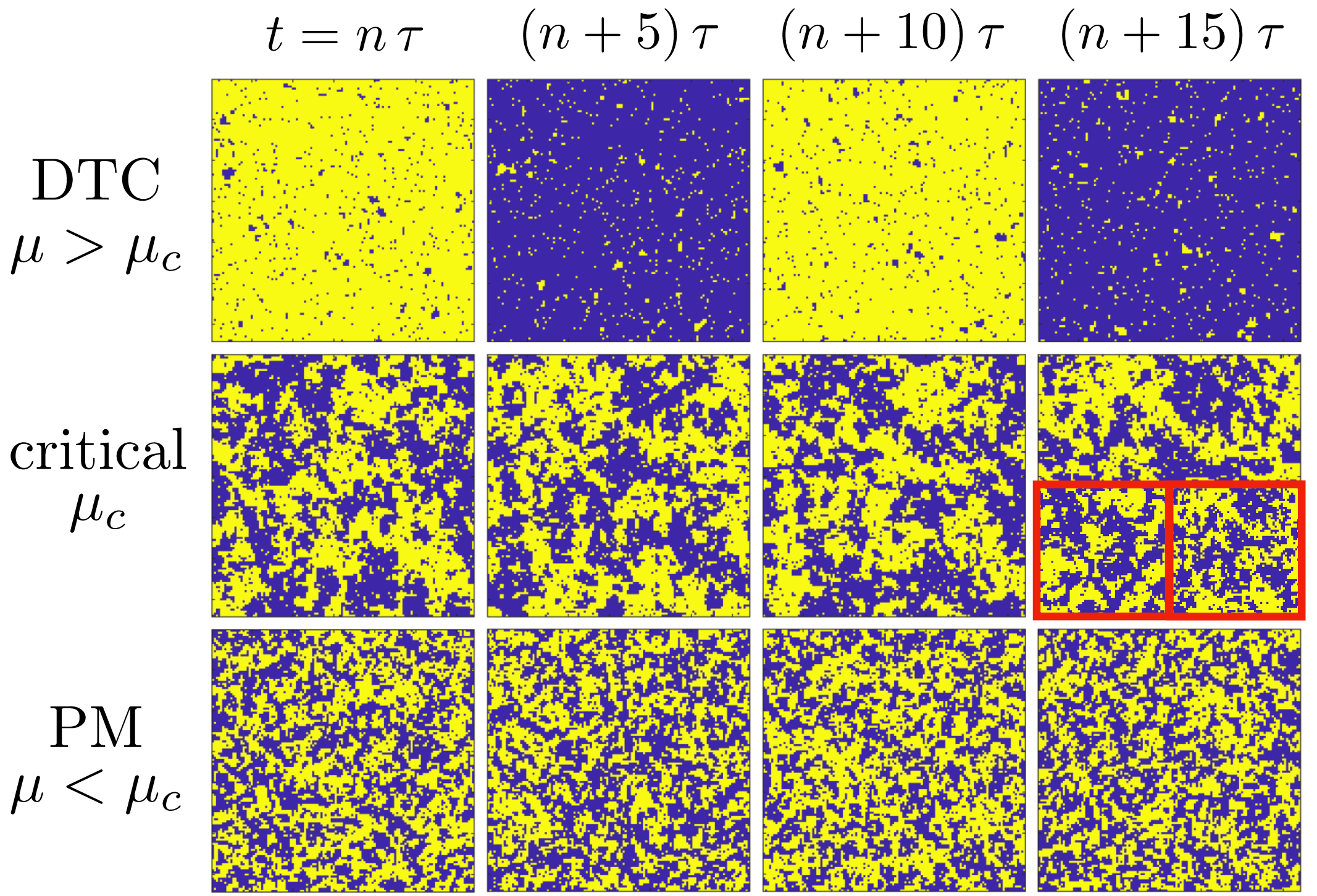}
	\caption{{\bf Stroboscopic trajectories.} Instantaneous configurations from representative trajectories under DTC conditions (top, $\mu=0.95>\mu_c$), near criticality (middle, $ \mu=0.9\approx\mu_c $), and for a paramagnetic NESS (bottom, $ \mu=0.8<\mu_c $). Here, $ \tau=1 $. We show a succession of snapshots for size $L=128 $ at times corresponding to odd and even cycles separated by $5 \tau$, with $t = \tau \times \{ 1985, 1990, 1995, 2000 \}$. The top row shows how the magnetization alternates in the DTC. The inset at the bottom of the last panel in the middle row compares a $64\times64$ (on the left) and a coarse-grained $128\times128$ (on the right) configurations on the same scale to illustrate the self-similarity of the system at $\mu_c$.}
	\label{fig:movies}
\end{figure}

Figure~\ref{fig:movies} shows stroboscopically sampled configurations along representative trajectories in the various regimes of the dynamics. The top row corresponds to the DTC, with $\mu > \mu_c$. Here, the magnetization at even and odd number of cycles alternates, displaying subharmonic behavior. The bottom row corresponds to the paramagnetic stroboscopic NESS at $\mu < \mu_c$. Beyond fluctuations, there is no distinction between even and odd cycles, and discrete time symmetry remains unbroken. The middle row corresponds to conditions near criticality, $\mu \approx \mu_c$. Here, the presence of a continuous phase transition should result in scale invariance~\cite{Chaikin:1995}. Indeed, this can be seen in the Inset to the rightmost panel of the middle row, in which we consider two typical configurations of the system for two different sizes, $ L=64 $ and $ L=128 $, respectively. The latter spin configuration is re-scaled through a majority voting coarse graining on $ 2\times2 $ plaquettes. Clearly, the $ L=64 $ and the re-scaled $ L=128$ configurations show an almost perfect self-similarity, thus confirming scale invariance at criticality.

\begin{figure}[t]
	\centering
	\includegraphics[width=\columnwidth]{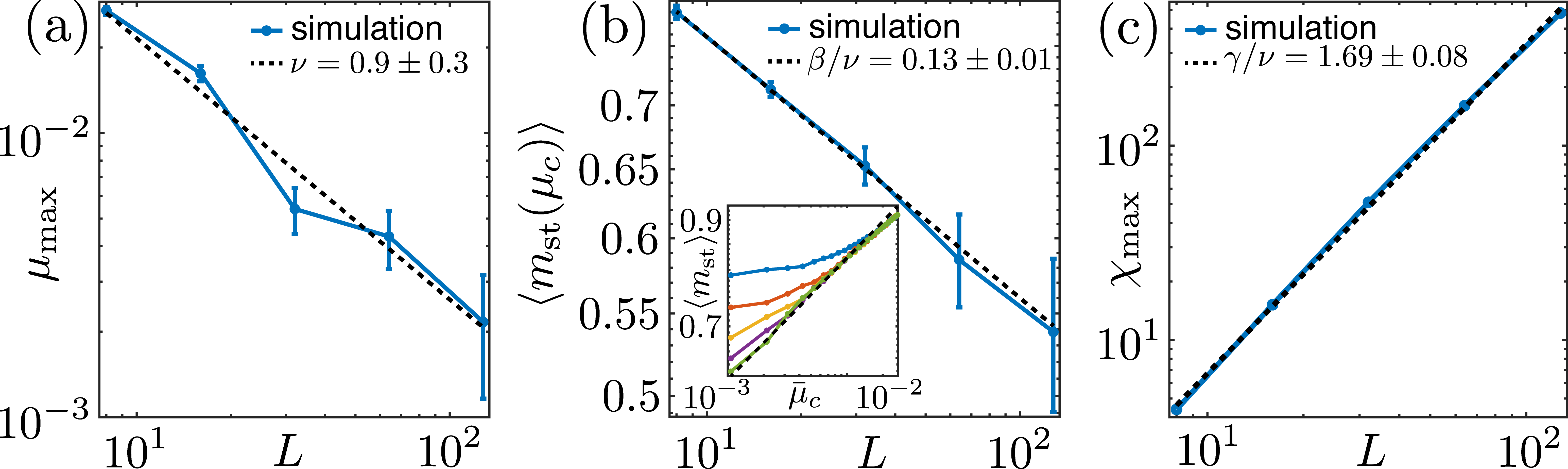}
	\caption{{\bf Critical exponents of the DTC transition obtained via finite size scaling.}
		(a) Plot of $ \mu_\mathrm{max} \sim L^{-1/\nu} $, the point in which $ \chi $ takes its maximum value $\chi_\mathrm{max} $ as a function $ L $. The fit results in $ \nu=0.9\pm0.3$.
		(b) Plot of $ \av{m_\mathrm{st}(\mu_c)} \sim L^{-\beta/\nu}$. The fit gives $ \beta/\nu=0.13\pm0.01 $. Using the value of $ \nu $ obtained in panel (a) we get $ \beta=0.12\pm0.04 $. Inset: plot of $ \av{m_\mathrm{st}} $ as a function of $ \bar{\mu}_c=(\mu-\mu_c)/\mu_c $ for the same sizes as in Fig.~\ref{fig:magandPD}(c). The slope $ \beta $ of the fitting curve (black, dashed) has been fixed at the value obtained from the main panel. 
		(c) Maximum value of the susceptibility, $ \chi_\mathrm{max}\sim L^{\gamma/\nu} $, as a function of $ L $. The fit leads to $ \gamma/\nu=1.69\pm0.08 $. Using the result for $ \nu $ from panel (a) we obtain $ \gamma=1.8\pm0.6 $. In panels (b, c), $ \mu_c $ is extracted as explained in Fig.~\ref{fig:magandPD}.}
	\label{fig:exp}
\end{figure}

\bigskip
\noindent
{\bf \em Critical properties of the DTC transition.--} Results from previous sections are indicative of a continuous phase transition with an emerging divergent correlation length at criticality. The latter describes the behavior of the (equal-time) connected correlation function,
\begin{equation}\label{eq:Cc}
C_{i,j}(t) \equiv \av{\sigma_i(t)\sigma_j(t)}-\av{\sigma_i(t)}\av{\sigma_j(t)},
\end{equation} 
at large (spatial) distances. In this limit, we expect that $ C_{i,j}\sim e^{-r_{ij}/\xi} $, with $ \xi $ being the correlation length and $ r_{ij} \equiv |\bm{r}_i-\bm{r}_j| $ the distance between spins $i$ and $j$. In a continuous phase transition a power-law behavior emerges in various thermodynamic quantities as the critical point is approached~\cite{Chaikin:1995}. Examples for such quantities include $ \xi $, the order parameter $ \av{m_{\mathrm{st}}} $ and its susceptibility $ \chi $, defined as
\begin{equation}\label{eq:chi}
\chi \equiv L^2\left(\av{m_{\rm st}^2}-\av{m_{\rm st}}^2\right).
\end{equation}
From the theory of continuous phase transitions, the following critical scaling relations are expected to hold in the stroboscopic NESS near the critical point
\begin{equation}
\av{m_{\rm st}} \sim |\mu-\mu_c|^\beta , 
\;\;
\xi \sim |\mu-\mu_c|^{-\nu}, 
\;\;
\chi \sim |\mu-\mu_c|^{-\gamma},
\label{eq:scaling}
\end{equation}
with $ \beta $, $ \nu $, and $ \gamma $ being the critical exponents corresponding to the various thermodynamic quantities. Despite Eq.~\eqref{eq:scaling} strictly holds in the thermodynamic limit, critical exponents can be efficiently extracted through a finite-size scaling analysis~\cite{Newman:1999,Binder:2010,Sandvik:2010}. In a finite system the correlation length $ \xi $ is bounded by the system size $ L $ and all the various thermodynamic quantities (e.g., $ \chi $) saturate when $ \xi\sim L $. Close to the critical point, by using that $ |\mu-\mu_c|\sim L^{-1/\nu} $ [see the equation in the middle of Eq.~\eqref{eq:scaling}], the following scaling relations can be obtained~\cite{Newman:1999,Sandvik:2010} 
\begin{equation}\label{eq:finite_size}
\mu_\mathrm{max} \sim L^{-1/\nu}, 
\; 
\chi_\mathrm{max}\sim L^{\gamma/\nu},
\;
\av{m_\mathrm{st}(\mu_c)}\sim L^{-\beta/\nu}.
\end{equation}
For a given $ L $, $ \mu_\mathrm{max} $ corresponds to the value of $ \mu $ in which $ \chi $ takes its maximum value, $ \chi(\mu_\mathrm{max}) = \chi_{\mathrm{max}} $, and $ \av{m_\mathrm{st}(\mu_c)} $ is the value of the staggered magnetization at the critical point. From Eq.~\eqref{eq:finite_size}, by inspecting different system sizes, it is possible to extract the values of $ \nu $, $ \beta/\nu $, and $ \gamma/\nu $. The behavior of $ \mu_\mathrm{max}$, $ \chi_\mathrm{max} $, and $ \av{m_\mathrm{st}(\mu_c)} $ as a function of $ L $ is shown in Fig.~\ref{fig:exp}. By fitting with the power laws in Eq.~\eqref{eq:finite_size}, we obtain the following values for the critical exponents of the disorder-DTC phase diagram: $ \nu=0.9\pm0.3 $, $ \beta=0.12\pm0.04 $ and $ \gamma=1.8\pm0.6 $. These values are compatible with the ones of the equilibrium 2D Ising model~\cite{Chaikin:1995},  $(\nu_{2D},\beta_{2D},\gamma_{2D}) = (1,1/8,7/4)$, even though in our case they correspond to a non-equilibrium phase transition, as shown in next section. To benchmark the validity of the fitting procedure, in the Inset of Fig.~\ref{fig:exp}(b) we plot $ \av{m_\mathrm{st}} $ as a function of $ \mu_c=(\mu-\mu_c)/\mu_c>0 $ for different sizes and checked that, for large $ L $, it follows the behavior predicted in Eq.~\eqref{eq:scaling}, i.e., $ \av{m_\mathrm{st}}\sim |\mu-\mu_c|^\beta$. 

\begin{figure}[t]
	\centering
	\includegraphics[width=\columnwidth]{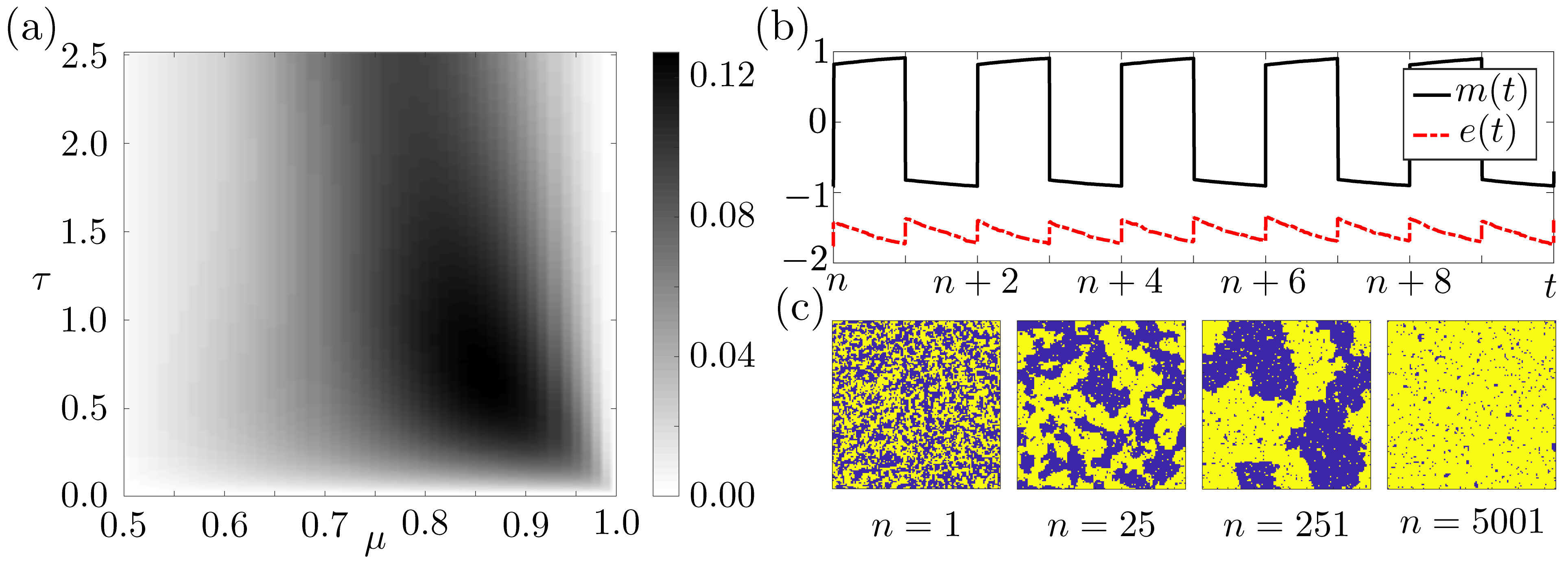}
	\caption{{\bf Non-equilibrium nature of the DTC.}
	(a) Sum of the absolute values of all elementary stationary currents $J_{\bm{\sigma} \rightarrow \bm{\sigma}'}$ for different values of the parameters $ (\tau,\mu) $ for $ L=3 $. A non-zero value of this witnesses the non-equilibrium nature of the Floquet dynamics. (b) Time-dependent magnetization $m(t)$ and energy $e(t)=1/L^2\sum_{\av{i,j}} \sigma_i(t) \sigma_j(t)$ for a system with $L=128$. The value of $n$ is chosen large enough so that the curves look stationary (in this case $n=100$ is sufficient). The behavior of both $m(t)$ and of $e(t)$, which is not time-reversible, makes it evident that the process generating these trajectories has a non-equilibrium character. (c) Coarsening dynamics in the instantaneous configurations of the system for $ t=\tau\times\{ 1, 25, 51, 5001 \} $ with $ m_{\mathrm{in}}=0.5 $. In panels (b) and (c), $ \tau=1$ and $\mu=0.95$.}
	\label{fig:noneq}
\end{figure}

\bigskip
\noindent
{\bf \em Non-equilibrium nature of the DTC phase.---} To assess the non-equilibrium nature of the DTC phase transition, we now analyze the properties of the Floquet evolution operator $ \mathcal{F}_\tau $. At stationarity we have $ \mathcal{F}_\tau \ket{\rho_\mathrm{NESS}}=\ket{\rho_\mathrm{NESS}} $, with $ \ket{\rho_\mathrm{NESS}} $ the system's stationary state. Here, the NESS probability of finding the system in a given spin configuration $ \bm{\sigma} $ is $ \rho_{\bm{\sigma}}=\braket{\bm{\sigma}}{\rho_{\mathrm{NESS}}} $, while the transition rate from the configuration $ \bm{\sigma} $ to $ \bm{\sigma'} $ is $ \mathcal{T}_{\bm{\sigma}\rightarrow\bm{\sigma}'}=\left< \bm{\sigma}'\right| \mathcal{F}_\tau \left| \bm{\sigma} \right>$. In these discrete time dynamics, one can define the average {\em elementary currents} between two configurations as $ J_{{\bm{\sigma}} \to {\bm{\sigma}}'} \equiv \rho_{{\bm{\sigma}}} \mathcal{T}_{{\bm{\sigma}} \to {\bm{\sigma}}'} - \rho_{{\bm{\sigma}}'} \mathcal{T}_{{\bm{\sigma}}' \to {\bm{\sigma}}}$. 
In equilibrium settings, detailed balance requires $ \rho_{{\bm{\sigma}}} \mathcal{T}_{{\bm{\sigma}} \to {\bm{\sigma}}'} = \rho_{{\bm{\sigma}}'} \mathcal{T}_{{\bm{\sigma}}' \to {\bm{\sigma}}} $ and, thus, on average no currents can be observed between any pairs of spin configurations ${\bm{\sigma}}$ and ${\bm{\sigma}}'$, i.e., $ J_{{\bm{\sigma}} \to {\bm{\sigma}}'}=0\ \forall\bm{\sigma},\bm{\sigma'} $. As such, in order to prove the non-equilibrium character of the phase transition that we are investigating, it would be sufficient to show the presence of at least one non-zero current \cite{Seifert2012}. For small sizes, the NESS $ \ket{\rho_{\mathrm{NESS}}} $ can be found through exact diagonalization of $\mathcal{F}_\tau$, while he probabilities $ \rho_{\bm{\sigma}} $ and the transition rates $ \mathcal{T}_{{\bm{\sigma}} \to {\bm{\sigma}}'} $ can be directly evaluated by listing all the possible spin configurations. One can then construct the matrix current, whose entries are the elementary currents between all the possible spin configurations $ J_{{\bm{\sigma}} \to {\bm{\sigma}}'} $, and evaluate its $ L_1 $-norm (defined as $\sum_{{\bm{\sigma}},{\bm{\sigma}}'} | J_{{\bm{\sigma}} \to {\bm{\sigma}}'}|/2$). Figure~\ref{fig:noneq}(a) gives the $ L_1 $-norm of the current matrix for a system with $ L=3 $ and for a range of values of $(\tau,\mu)$ and shows that, except for the limits $\tau=0$ and $\mu = 1$, there exists at least one non-vanishing average current, resulting in a violation of detailed balance and, therefore, demonstrating that the disorder-DTC order is a non-equilibrium phase transition.

Such a non-equilibrium nature of the dynamics is also clearly apparent in continuous-time resolved dynamical realizations, as displayed in Fig. \ref{fig:noneq}(b) for the energy and the magnetization, which are manifestly not time-reversible.

Having shown that the asymptotic state is a non-equilibrium one, we can ask whether the analogy with the 2D Ising model extends beyond the stroboscopic NESS. A simple test is to consider the dynamics starting from an initial state with zero magnetization (such as a quench from a random configuration). Figure~\ref{fig:noneq}(c) shows that under DTC conditions the stroboscopic dynamics exhibits {\em coarsening}, displaying progressive growth of domains of {\em both} magnetized states (before eventual collapse to one of the two in the last panel, which is a finite size effect). As the relaxation step of the protocol is standard Glauber dynamics at zero temperature we could have expected coarsening only within a period, but the fact that it survives for long times despite the periodic driving is non-trivial, and constitutes another indication that the DTC state is robust. 
 
\bigskip
\noindent
{\bf \em Conclusions.---} We have shown that a DTC phase can be obtained in a fully classical and thermal setting - and in the absence of disorder or any form of classical localization - as a symmetry breaking transition of a driven 2D Ising model. 
The classical nature of the problem we consider allowed us, by means of numerical simulations of large system sizes and finite-size scaling (something often outside the range of possibility in both closed and open quantum systems), to demonstrate convincingly that the transition to a DTC is indeed a phase transition. Our results suggest that when observed stroboscopically, this phase transition in the NESS is in the 2D Ising universality class. The mechanism for DTCs described here is simple and easily generalizable. For example, we expect that driven Potts models will lead to DTCs with periods larger than two~\cite{Surace:2019,Pizzi:2019}. The simplicity of the scheme also suggests that it should be possible to observe these classical DTCs in experiments such as magnetic solid state systems and other platforms with order-disorder transitions with a scalar order parameter. In particular, the dynamics we have studied in this work can be implemented in a setup of dissipative Rydberg atoms~\cite{Low:2012,Lee:2012,Marcuzzi:2014} in the limit of strong dephasing~\cite{Lesanovsky:2013,Marcuzzi:2014b,Levi:2016} and driven by a periodic sequence of slightly imperfect $ \pi $-pulse rotations, which simulates the random inversion of a fraction of spins.
\bigskip

\begin{acknowledgments}
The research leading to these results has received funding from the  European  Research Council under the European Union's Seventh Framework Programme (FP/2007-2013)/ERC Grant Agreement No.~335266 (ESCQUMA), and from EPSRC Grants No.~EP/R04421X/1 and EP/N03404X/1. I.L.~gratefully acknowledges funding through the Royal Society Wolfson Research Merit Award. The simulations used resources provided by the University of Nottingham High-Performance Computing Service.
\end{acknowledgments}

\footnotetext[1]{Ising models driven via a periodically oscillating magnetic field has been studied in Refs.~\cite{Chakrabarti1999,Korniss2000,Park2013}. However, they do not lead to DTC behavior. Note that the inversion step of Eq.~\eqref{eq:Floquet} in our protocol cannot be realized by a time-dependent change of the energy function (such as a with an oscillatory magnetic field).}

\bibliography{bibliography.bib}

\end{document}